\documentclass[aps,prb,preprint,showpacs,superscriptaddress]{revtex4}
\usepackage{graphicx}
\usepackage{amsmath}
\begin{document}
\newcommand\hata{{\hat{a}}}
\newcommand\hatc{{\hat{c}}}
\newcommand\hatd{{\hat{d}}}
\newcommand\hatn{{\hat{n}}}
\newcommand\hatH{{\widehat{H}}}
\newcommand\hatI{{\widehat{I}}}
\newcommand\hatN{{\widehat{N}}}
\newcommand\tilc{{\tilde{c}}}
\newcommand\tild{{\tilde{d}}}
\newcommand\tilk{{\tilde{k}}}
\newcommand\tiln{{\tilde{n}}}
\newcommand\tilE{{\widetilde{E}}}
\newcommand\tilH{{\widetilde{H}}}
\newcommand\tilI{{\widetilde{I}}}
\newcommand\tilN{{\widetilde{N}}}
\newcommand\tilP{{\widetilde{P}}}
\newcommand\tilV{{\widetilde{V}}}
\newcommand\tilS{{\widetilde{S}}}
\newcommand\tilT{{\widetilde{T}}}
\newcommand\tilU{{\widetilde{U}}}
\newcommand\tilW{{\widetilde{W}}}
\newcommand\varF{{\mathcal{F}}}
\newcommand\ket[1]{|{\textstyle{#1}}\rangle}
\newcommand\bra[1]{\langle{\textstyle{#1}}|}
\newcommand\avg[1]{\langle{\textstyle{#1}}\rangle}
\let\up=\uparrow
\let\down=\downarrow
\title{Pair Tunneling and shot noise through a single molecule in a
  strong electron-phonon coupling}

\author{Myung-Joong Hwang}%
\affiliation{Department of Physics, Pohang University of Science and
  Technology, Pohang 790-784, Korea}%
\author{Mahn-Soo Choi}%
\email{choims@korea.ac.kr}%
\affiliation{Department of Physics, Korea University, Seoul 136-713,
  Korea}%
\affiliation{Department de F\'isica, Universitat de les Illes Balears,
  E-07122 Palma de Mallorca, Spain}%
\author{Rosa L\'opez}%
\affiliation{Department de F\'isica, Universitat de les Illes Balears,
  E-07122 Palma de Mallorca, Spain}%

\begin{abstract}
We investigate the electronic transport through a single molecule in a
strong electron-phonon coupling regime.  Based on a particle-hole
transformation which is made suitable for non-equilibrium situation, we
treat the pair tunneling and cotunneling on an equal footing.  We
propose an experimental setup to enhance the visibility of pair
tunneling, which has no Franck-Condon suppression. We also
discuss the shot noise characteristics.
\end{abstract}
\maketitle

\section{Introduction} 
Many transport features of single molecules are ascribed to the
coupling between electronic and vibrational
modes~\cite{Cuniberti05a,Park00a}. 
Especially, the electron-phonon coupling makes the effective charging
energy $U$ reduced and even negative.
The negative $U$ makes the double occupancy of the molecular level
energetically favorable, and allows the tunneling of pairs. Although
most studies on the negative $U$ have focused on the Kondo
regime~\cite{kondo,Cornaglia04a,Cornaglia05a}, a recent
work~\cite{Koch06a} pointed out that the pair tunneling can
significantly affect transport above the Kondo temperature ($T_K$) as
well, and leads to several very unusual features~\cite{Koch06a}. In
particular, the resonance width of the pair tunneling conductance was
only determined either by the temperature (linear) or by the bias
voltage (non-linear); but not by the lead-molecule hybridization as in
normal sequential tunneling of single particles.  Further, the pair
tunneling conductance was exactly twice the normal cotunneling
contribution.

However, we note that there are several physically important
questions that have not been addressed yet.  Firstly, there is no
justification why the pair tunneling be treated separately in a
sequential tunneling picture apart from the normal cotunneling
processes. In this work, we use a different approach and treat both
on an equal footing. 
Secondly, it is a general observation that the pair tunneling is
usually subject to the exponential Franck-Condon suppression. Then, how can we
observe the pair tunneling processes experimentally in a broad
background of normal cotunneling contributions? Here we
propose an experimental setup enabling one to observe the pair
tunneling. 
Thirdly, is there a simple physical explanation why the pair
tunneling conductance is exactly twice the normal cotunneling
contribution? Can one attribute it to the double charge of the pair
(like in a tightly bound pair objects)? We find that this is not the
case.

In addition to the previous questions on average current
(conductance), for a better understanding of the nature of the pair
tunneling, one can also investigate the fluctuations of the current.
In many systems, the effective charge of the elementary excitations
has been identified by the shot noise
characteristics~\cite{Blanter00a,ChoiMS03a,ChoiMS01d}. Interesting
questions would then be: Should the pair tunneling events give the
fluctuations corresponding to the Fano factor equal to 2? How
different are the ¡°pairs¡± tunneling through the device from
tightly bound pair objects?
In this work, we address all these questions and provide further
physical insights to the pair-tunneling transport (above $T_K$),
based on particle-hole transformation.

The paper is organized as following: In Section~\ref{sec:particle-hole}
we establish the particle-hole transformation, which enables us to treat
the pair-tunneling processes on the same footing as the normal
single-electron cotunneling processes.  This transformation, augmented
with the exchange of the electrode indices, is suitable for
non-equilibrium (as well as equilibrium) transport.  In
Section~\ref{sec:cotunneling-rates} we evaluate all the relevant
cotunneling rates based on the particle-hole transformation.  This will
make clear the difference between the underlying physical processes
governing the pair tunneling and normal single electron cotunneling.
Moreover, we clearly point out how the pair tunneling transport is
exponentially suppressed by the Franck-Condon effect in realistic
molecular devices, which was overlooked in the previous
work.\cite{Koch06a} Section~\ref{sec:current} is then devoted to
reproduce the results on average current (and conductance) of previous
work to demonstrate the efficiency of our method.  The current
fluctuation noise and investigate the contribution to it from the pair
tunneling processes in Section~\ref{sec:noise}.  In
Section~\ref{sec:ccdqd} a realistic experiment is proposed where one can
observe the pair tunneling physics without suffering from the
Franck-Condon suppression, and provide the detailed analysis of the
setup.  Section~\ref{sec:conclusion} concludes the paper.

\section{Particle-hole transformation}
\label{sec:particle-hole}

We describe the molecular device with an Anderson-Holstein model
(AHM), whose Hamiltonian has the form
$\hatH=\hatH_L+\hatH_R+\hatH_M+\hatH_T$. The left ($L$) and right
($R$) electrodes are described by the non-interacting electrons
\begin{equation}
\hatH_\ell = \sum_{k\sigma}(\epsilon_{k}-eV_\ell) \hatc_{\ell
k\sigma}^\dag\hatc_{\ell k\sigma} \quad (\ell=L,R)
\end{equation}
Here $\hatc_{\ell k\sigma}^\dag$ creates an electron with momentum
$k$ and spin $\sigma$ on the lead $\ell=\{L,R\}$. The molecule has
both the electronic degrees of freedom, described by the fermion
operators $\hatd$ and $\hatd_\sigma^\dag$
($\hatn_\sigma=\hatd_\sigma^\dag\hatd_\sigma$,
$\hatn=\hatn_\up+\hatn_\down$), and the vibrational mode with
frequency $\omega_0$, described by the boson operators
$\hata$ and $\hata^\dag$:
\begin{equation}
\hatH_M = \sum_\sigma\epsilon_0\hatd_\sigma^\dag\hatd_\sigma +
U_0\hatn_\up\hatn_\down - \lambda\hbar\omega_0(\hatn-1)(\hata^\dag +
\hata) + \hbar\omega_0\hata^\dag\hata \,,
\end{equation}
where $\lambda$ is the dimensionless electron-phonon coupling constant
between the two degrees of freedom. The electron tunneling between the
each lead and the molecule is expressed as
\begin{equation}
\hatH_T = \sum_{\ell k\sigma} \left(T_{\ell k}\hatc_{\ell
k\sigma}^\dag\hatd_\sigma + h.c.\right) \,.
\end{equation}
The molecule-electrode couplings are characterized by the
hybridization parameters
\begin{math}
\Gamma_{\ell} \equiv 2\pi\sum_k|T_{\ell
k}|^2\delta(E-\epsilon_{k\sigma})
\end{math}.  As usual, we ignore the weak energy dependence of
$\Gamma_\ell$.  We also put $\Gamma=\Gamma_L+\Gamma_R$.

\begin{figure}
\centering
\includegraphics*[width=7cm]{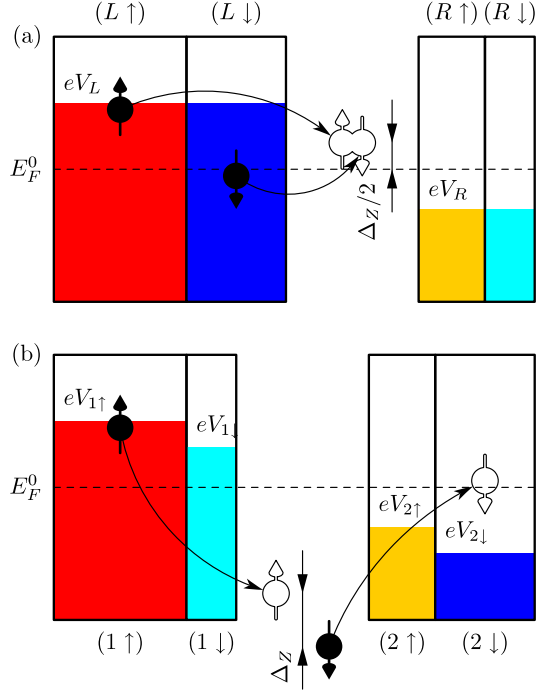}
\caption{(color on-line) An illustration of the particle-hole
  transformation.  A pair tunneling in the original model (a)
  corresponds to a spin-flip cotunneling in the equivalent model (b).
  To emphasize the exchange of the electrodes in the particle-hole
  transformation, different widths (colors) have been used for electrode
  bands.  Spin-preserving cotunneling processes (not shown) remain the
  same in both models. In (a), shown on the molecule site is the
  \emph{single-particle} energy level for the double occupancy.}
\label{pair::fig:1}
\end{figure}

The molecular part $H_M$ is diagonalized as
\begin{equation}
H_M = e^{+S}\left[
  \sum_\sigma\epsilon_d d_\sigma^\dag d_\sigma
  + Un_\up n_\down\right]e^{-S} + \hbar\omega_0a^\dag a
\end{equation}
by the canonical transformation
\begin{math}
S = \lambda (n-1)(a^\dag - a)
\end{math}.\cite{Schuttler88a,Cornaglia04a} The renormalized molecular
level $\epsilon_d$ and on-site interaction $U$ are given by
\begin{math}
\epsilon_d = \epsilon_0 + \lambda^2\hbar\omega_0
\end{math}
and
\begin{math}
U = U_0 - 2\lambda^2\hbar\omega_0
\end{math}, respectively. In the strong coupling limit
($\lambda^2\gg{}U_0/2\hbar\omega_0$) the on-site interaction $U$
becomes \emph{negative} so that a double occupancy of the
molecular level is energetically favored to a single occupancy. 
Indeed, by removing the phonon degrees of freedom with a Schrieffer-Wolf
(SW) transformation, it was shown that electrons are allowed to tunnel
\emph{in pairs} into the molecule~\cite{Koch06a}; see Fig.~\ref{pair::fig:1}
(a).
However, it should be stressed that the situation considered in
Ref.~\onlinecite{Koch06a} is rather special and in fact equivalent to a
negative-$U$ Anderson model (\emph{isotropic} Kondo model).  Earlier, it
was shown in Ref.~\onlinecite{Schuttler88a} that the model is, in
general, equivalent to an \emph{anisotropic} Kondo model (not equivalent
to a negative-$U$ Anderson model).\cite{endnote:1}
We will come back to this issue later.

It has been proved useful to map a negative-$U$ impurity model to an
equivalent model with positive interaction by a particle-hole
transformation (PHT)~\cite{pht}.  In the same spirit, we directly apply
the PHT to the AHM keeping the phonon modes~\cite{endnote:2}.
We first choose an one-to-one correspondence
\begin{math}
k \mapsto \tilk
\end{math}
such that
\begin{math}
\epsilon_\tilk = -\epsilon_{k}
\end{math} (we assume symmetric conduction bands).  Following
Refs.~\onlinecite{Schuttler88a,pht}, we then make the PHT for down spins
\begin{equation}
\hatd_\down \mapsto \tild_\down^\dag \,,\quad
\hatc_{Lk\down} \mapsto \tilc_{2\tilk\down}^\dag \,,\quad
\hatc_{Rk\down} \mapsto \tilc_{1\tilk\down}^\dag \,,
\end{equation}
keeping the up spins unchanged
\begin{equation}
\hatd_\up \mapsto \tild_\up \,,\quad
\hatc_{Lk\up} \mapsto \tilc_{1k\up} \,,\quad
\hatc_{Rk\up} \mapsto \tilc_{2k\up} \,.
\end{equation}
It is emphasized that the lead indices for the down spins have been
exchanged ($L\to 2, R\to 1$) compared with those for the up spins ($L\to
1, R\to 2$).  This is a small yet important difference between the
mapping here and that in Refs.~\onlinecite{Schuttler88a,pht}; the
lead-index exchange is not necessary at equilibrium.  Under this
transformation, $H_L+H_R$ is transformed to $\tilH_1+\tilH_2$ with
\begin{equation}
\label{pair::eq:1}
\tilH_\ell =
\sum_{k\sigma}(\epsilon_{k\sigma}-e\tilV_{\ell\sigma}) \tilc_{\ell
k\sigma}^\dag \tilc_{\ell k\sigma} \,,
\end{equation}
and $H_T$ to
\begin{equation}
\label{pair::eq:2}
\tilH_T = \sum_{\ell k\sigma} \left(\tilT_{\ell
k\sigma} \tilc_{\ell k\sigma}^\dag \tild_\sigma
  + h.c.\right) \,.
\end{equation}
The effective bias voltages $e\tilV_{\ell\sigma}$ become, in general,
spin-dependent:
\begin{math}
e\tilV_{1\up} = eV_L
\end{math},
\begin{math}
e\tilV_{1\down} = -eV_R
\end{math},
\begin{math}
e\tilV_{2\up} = eV_R
\end{math}
\begin{math}
e\tilV_{2\down} = -eV_L
\end{math}.
For the symmetric bias, which is assumed here, they are spin-independent:
\begin{math}
e\tilV_{1\up} = e\tilV_{1\down} = eV_b/2
\end{math},
\begin{math}
e\tilV_{2\up} = e\tilV_{2\down} = -eV_b/2
\end{math}.  The tunneling amplitudes $\tilT_{\ell k\sigma}$ are also
spin-dependent and given by
\begin{equation}
\label{pair::eq:3}
\tilT_{1k\up} = T_{Lk\up} \,,\quad
\tilT_{1k\down} = -T_{R\tilk\down}^* \,,\
\tilT_{2k\up} = T_{Rk\up} \,,\quad \tilT_{2k\down} =
-T_{L\tilk\down}^* \,.
\end{equation}
Their spin-dependence disappears only for symmetric junctions
($\Gamma_L=\Gamma_R=\Gamma/2$):
\begin{math}
\tilde\Gamma_{\ell\sigma} = \Gamma/2
\end{math}
$(\ell=1,2)$.  Later, we will see that the junction asymmetry affects
significantly the transport properties of the device.
Finally, the molecular part now takes the form
\begin{equation}
e^{+\tilS}\left[
  \sum_\sigma\tilde\epsilon_{d\sigma}\tild_\sigma^\dag\tild_\sigma
  + \tilU\tiln_\up\tiln_\down
+ a^\dag a\right]e^{-\tilS}
\end{equation} with
\begin{math}
\tilS = \lambda (\tiln_\up-\tiln_\down)(a^\dag - a)
\end{math}.
The effective on-site interaction
\begin{math}
\tilU \equiv -U
\end{math}
thus becomes positive, and the molecular level
\begin{math}
\tilde\epsilon_{d\sigma} \equiv {U}/{2} + \sigma\Delta_Z/2
\end{math}
has an effective Zeeman splitting
\begin{equation}
\Delta_Z \equiv 2\epsilon_d + U = 2\epsilon_0 + U_0 \,.
\end{equation}
The eigenstates are given by
\begin{math}
\ket{\tilde0}\ket{m}
\end{math}
(with energy
\begin{math}
\tilE_{0,m} = m\hbar\omega_0
\end{math}),
\begin{math}
\ket{\tilde2}\ket{m}
\end{math}
(\begin{math}
\tilE_{2,m} = m\hbar\omega_0
\end{math}), and
\begin{math}
D(\sigma\lambda)\ket{\tilde\sigma}\ket{m}
\end{math}
(\begin{math}
\tilE_{\sigma,m} = m\hbar\omega_0 + \tilde\epsilon_{d\sigma}
\end{math}), where $\ket{\tilde0}$ (empty), $\ket{\tilde2}$ (doubly
occupied), and $\ket{\tilde\sigma}$ (singly occupied by spin $\sigma$)
are electronic states of the molecule, $\ket{m}$ is the phonon state,
and
\begin{math}
D(\alpha) = \exp\left[\alpha(a^\dag-a)\right]
\end{math}.
\cite{Schuttler88a,Cornaglia04a,Cornaglia05a}

In the strong coupling regime ($\lambda^2\gg U_0/2\hbar\omega_0$), the
equivalent model has
$\tilU\gg\Gamma$ and $\tilde\epsilon_{d\sigma}\ll-\Gamma<0$; the
so-called local-moment regime.  The transport thus occurs only through
cotunneling processes (above $T_K$).
The PHT maps the electronic state $\ket{0}$ of the molecule in the
original model to
$\ket{\tilde\down}\equiv\tilde{d}_\down^\dag\ket{\tilde{0}}$ in the
equivalent model; likewise, $\ket{\up}\mapsto\ket{\tilde2}$,
$\ket{\down}\mapsto\ket{\tilde0}$, and $\ket{2}\mapsto\ket{\tilde\up}$.
The usual cotunneling process in the original model thus corresponds to
the spin-preserving cotunneling (SPC) in the equivalent model, and the
pair tunneling [Fig.~\ref{pair::fig:2}(a)] to the spin-flip cotunneling (SFC)
[Fig.~\ref{pair::fig:2}(b)].
Namely, unlike in Ref.~\onlinecite{Koch06a}, in our picture all the
relevant processes are treated on an equal footing, in terms only of
``cotunneling''.  This will allow us to infer further insight into the
pair-tunneling transport based on the well-established theory of
cotunneling transport~\cite{Sukhorukov01a}.

A few further remarks are in order: 
(i) In most experiments, the two leads are identical.  But the mapping
applies to general cases, including ferromagnetic leads~\cite{Rosa}. 
(ii) The transformed Hamiltonian arises from a mathematical mapping.
Yet, the model itself is \emph{physical} (experimentally realizable).
Among others, an interesting realization will be the capacitively
coupled double quantum dot (CCDQD); see Fig.~\ref{pair::fig:4}.

\section{Cotunneling rates}
\label{sec:cotunneling-rates}

Both the average current and the shot
noise characteristics can be obtained from the rate equation
\begin{figure}
\centering
\includegraphics*[width=8cm]{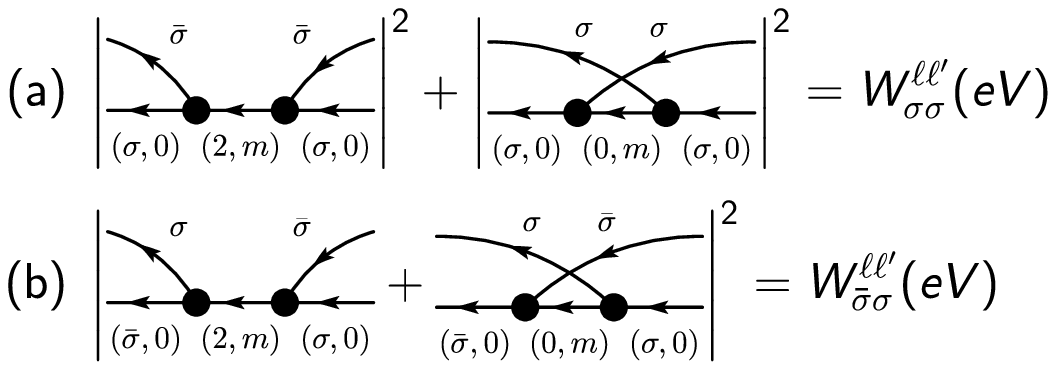}
\caption{Cotunneling rates in the equivalent model. (a) Spin-preserving
  cotunneling, corresponding to the usual single-particle cotunneling in
  the original model.  (b) Spin-flip cotunneling, corresponding to the
  pair tunneling in the original model.}
\label{pair::fig:2}
\end{figure}
\begin{multline}
\label{pair::eq:4}
\frac{d\tilP_\sigma(N)}{dt}
= -\sum_{\ell\ell',\sigma'}\tilW_{\sigma'\sigma}^{\ell'\ell}\tilP_\sigma(N)
+ \sum_{\ell}\tilW_{\sigma\bar\sigma}^{\ell\ell}\tilP_{\bar\sigma}(N)
\\\mbox{}
+ \sum_{\sigma'}\tilW_{\sigma\sigma'}^{21}\tilP_{\sigma'}(N-1)
+ \sum_{\sigma'}\tilW_{\sigma\sigma'}^{12}\tilP_{\sigma'}(N+1).
\end{multline}
Here $\tilP_\sigma(N,t)$ is the probability to find the molecule in the
electronic state $\ket{\tilde\sigma}$ and $N$ electrons in the lead
$2$ at time $t$.  $\tilW_{\sigma\sigma'}^{\ell\ell'}$ is the the rate of
the cotunneling process where an electron is transferred from lead $\ell'$
to $\ell$ and, at the same time, the \emph{molecular state} changes from
$\ket{\tilde\sigma'}$ to $\ket{\tilde\sigma}$. 
The cotunneling rates for the spin-preserving and spin-flip processes
are given respectively by
\begin{equation}
\label{pair::eq:5}
\tilW_{\sigma\sigma}^{\ell\ell'}
= \varF_{+}^2(u,\lambda)
\left[
  \frac{\tilde\Gamma_{\ell\bar\sigma}\tilde\Gamma_{\ell'\bar\sigma}+\tilde\Gamma_{\ell\sigma}\tilde\Gamma_{\ell'\sigma}}{2\pi\hbar\tilde\epsilon_{d\sigma}^2}
\right]F(e\tilV_{\ell\sigma}-e\tilV_{\ell'\sigma})
\end{equation}
\begin{equation}
\tilW_{\bar\sigma\sigma}^{\ell\ell'} = \varF_{-}^2(u,\lambda)
\tilde\Gamma_{\ell\sigma}\tilde\Gamma_{\ell'\bar\sigma} \left[
  \frac{1}{\tilde\epsilon_{d\bar\sigma}}
  + \frac{1}{\tilde\epsilon_{d\sigma}}
\right]^2\\\mbox{}\times
\frac{F(\tilde\epsilon_{d\bar\sigma}-\tilde\epsilon_{d\sigma}
+ e\tilV_{\ell\sigma}-e\tilV_{\ell'\bar\sigma})}{2\pi\hbar}
\end{equation}
where
\begin{math}
u \equiv -U/2\hbar\omega_0
\end{math}
and
\begin{math}
F(E) \equiv E/(e^{\beta E} - 1)
\end{math}.
One can immediately notice the differences between the two rates (see
also Ref.~\onlinecite{Lehmann06a}).  (i) As illustrated diagrammatically in
Fig.~\ref{pair::fig:2},\cite{Hewson93a} the SPC rate is an \emph{incoherent}
(classical) sum of two contributions (because the spins of the
conduction electrons are different between the two corresponding
diagrams) while the SFC rate is a \emph{coherent} sum of two
\emph{indistinguishable} contributions.  The latter gives rise to a
constructive interference for the SFC rate, and hence
$\tilW_{\bar\sigma\sigma}^{\ell\ell'}$ becomes exactly twice
$\tilW_{\sigma\sigma}^{\ell\ell'}$ at $eV_b\approx 0$ and $\Delta_Z\approx 0$.
This explains why the pair-tunneling conductance is twice the normal
cotunneling contribution.  (ii) They have different Franck-Condon (FC)
factors
\begin{equation}
\varF_\pm(u,\lambda) \equiv
e^{-\lambda^2}\sum_{m=0}^\infty
\frac{(\pm\lambda^2)^m}{m!}\frac{u}{m+u}
\end{equation}

The FC suppression arises at two different levels.  According to the FC
principle~\cite{FC}, each virtual tunneling amplitude itself has the FC
suppression factor:
\begin{math}
\bra{\tilde\sigma,0}H_T\ket{\tilde\alpha,m}\sim e^{-\lambda^2/2}
\end{math}
and
\begin{math}
\bra{\tilde\alpha,m}H_T\ket{\tilde\sigma,0} \sim e^{-\lambda^2/2}
\end{math}
($\alpha=0,2$).  For the \emph{overall} FC factor, however, the
contributions from various intermediate states should be all summed up.
(i) When $\hbar\omega_0\gg -U/2$ ($u\ll 1$),
$\ket{\tilde\alpha,m=0}$ is the only
intermediate state contributing to the overall cotunneling amplitude.
In this case, the overall FC factor is $e^{-\lambda^2}$ for
\emph{both} $\tilW_{\sigma\sigma}^{\ell\ell'}$ and
$\tilW_{\bar\sigma\sigma}^{\ell\ell'}$.  This is the limit considered in
Ref.~\onlinecite{Koch06a}, corresponding to the \emph{isotropic} Kondo model
(negative-U Anderson model).
(ii) However, if $\hbar\omega_0\ll -U/2$ ($u\gg 1$), all the intermediate
states $\ket{\tilde\alpha,m}$ with higher vibrational energies give
finite contributions.
The sum of these contributions just amounts to cancel the
FC suppression factor from the \emph{individual} virtual tunneling for
$\tilW_{\sigma\sigma}^{\ell\ell'}$, while it is not the case for
$\tilW_{\bar\sigma\sigma}^{\ell\ell'}$.  Consequently, the SFC
(the pair tunneling) is
exponentially suppressed compared with the SPC.
This is the case examined by most authors, including
Refs.~\onlinecite{Schuttler88a} and \onlinecite{Cornaglia04a}.

We define the \emph{relative} FC factor by
$\gamma\equiv\varF_{-}(u,\lambda)/\varF_{+}(u,\lambda)$.  In typical
experiments with C$_{60}$ molecule~\cite{Park00a}, $U_0\sim 300$ meV and
$\hbar\omega_0\sim 5$ meV.  Therefore, the condition for the case (i)
above is hardly satisfied, and $\gamma$ can be significantly (even
though not exponentially) smaller than $1$, and it may be difficult to
observe the pair tunneling.  Below (Section~\ref{sec:ccdqd}) we will
propose an experiment where the pair tunneling is not subject to the
Franck-Condon suppression.

\section{Current}
\label{sec:current}

The average dc current is given by
$I=\sum_{\sigma}\avg{\hatI_{d\sigma}}$ with
\begin{math}
\hatI_{d\sigma}
= (\hatI_{R\sigma} - \hatI_{L\sigma})/2
= (\tilI_{2\sigma} - \tilI_{1\sigma})/2
\end{math}, where
\begin{math}
\hatI_{\ell\sigma} = e{d\hatN_{\ell\sigma}}/{dt}
\end{math}
and
\begin{math}
\tilI_{\ell\sigma} = e{d\tilN_{\ell\sigma}}/{dt}
\end{math}.  Referring to the PHT (Fig.\ref{pair::fig:1}), one can
identify the two contributions to the total current $I=e(J_c+J_p)$,
namely,
\begin{equation}
\label{pair::eq:6} J_c = (J_{\up\up}+J_{\down\down})/2
\end{equation}
from the usual cotunneling and
\begin{equation}
\label{pair::eq:7} J_p = J_{\up\down}\tilP_\down +
J_{\down\up}\tilP_\up
\end{equation}
from the pair tunneling.  Here
\begin{math}
J_{\sigma\sigma'} \equiv \tilW_{\sigma\sigma'}^{21} -
\tilW_{\sigma\sigma'}^{12}
\end{math},
and
\begin{math}
\tilP_\sigma \equiv \sum_N\tilP_\sigma(N,t=\infty)
\end{math}
is the stationary probability distribution of the molecular state.
Putting the cotunneling rates in Eq.~(\ref{pair::eq:5}), one can easily
reproduce all the results in Ref.~\onlinecite{Koch06a}, in particular,
the linear conductance (up to the FC factor $\gamma^2$)
\begin{equation}
\label{pair::eq:8}
G = \frac{2e^2\Gamma_L\Gamma_R}{h}
\left[
  \frac{\gamma^2U^2}{\epsilon_d^2(\epsilon_d+U)^2}
  \frac{\beta(2\epsilon_d+U)}{2\sinh[\beta(2\epsilon_d+U)]}
  + \frac{1-f(2\epsilon_d+U)}{\epsilon_d^2}
  + \frac{f(2\epsilon_d+U)}{(\epsilon_d+U)^2}
\right]
\end{equation}
Here we just clarify the questions raised at the beginning.  In the
original sequential-tunneling treatment of the pair tunneling in
Ref.~\onlinecite{Koch06a}, it is not clear why and to what extent the
coherence between the subsequent pair-tunneling events can be ignored.
In the present picture, the pair tunneling and normal cotunneling are
treated on an equal footing, all in the cotunneling picture.
Thus, in order to go beyond the sequential-tunneling treatment of the
pair tunneling, viz., to the purely coherent resonant tunneling of
pairs, which leads to the Kondo effect~\cite{Cornaglia04a}, one has only
to go to higher orders.
Further, pair-tunneling contribution is exactly twice (for $\gamma=1$)
the usual cotunneling contribution due to the interference in the SFC
process (see above and Fig.~\ref{pair::fig:2}).

\section{Noise}
\label{sec:noise}

The current noise spectral density is given by
(at sufficiently low frequencies $\omega\ll-U/2$)
\begin{equation}
S(\omega) = \int_{-\infty}^\infty{d\tau}\; e^{+i\omega\tau}
\avg{\delta \hatI_d(\tau)\delta \hatI_d(0)
  +\delta\hatI_d(0)\delta\hatI_d(\tau)}
\end{equation}
where $\delta \hatI_d\equiv \hatI_d - \avg{\hatI_d}$.  The Fano
factor, defined by $S(0)/2eI$, is a representative characteristic of
the shot noise ($eV_b\gg k_BT$).  It may reveal not only the
correlated transport but also the effective charge of the
carriers~\cite{Blanter00a,ChoiMS03a,ChoiMS01d}. Using the quantum regression
theorem~\cite{Gardiner00a,ChoiMS03a,ChoiMS01d,shot} and the rate
equation~(\ref{pair::eq:4}), we obtain current-current correlation
function ($k_BT\ll eV_b\ll -U/2$ and $\Delta_Z\ll-U/2$)
\begin{equation}
\label{pair::eq:9} S(\omega) = 2eI
-4e^2\left[\frac{J_{\up\down}-J_{\down\up}}{\tilW_p}\right]
\left[\frac{J_{\up\down}\tilP_\down^2 - J_{\down\up}\tilP_\up^2}
  {1+\omega^2/\tilW_p^2}
\right]
\end{equation}
where
\begin{math}
\tilW_p = \sum_{\ell\ell'\sigma}\tilW_{\sigma\bar\sigma}^{\ell\ell'}
\end{math}.
The Fano factor thus has the form
\begin{equation}
\label{pair::eq:10}
\frac{S(0)}{2eI} = 1
- 2\left[\frac{J_{\up\down}-J_{\down\up}}{\tilW_p}\right]
\left[\frac{J_{\up\down}\tilP_\down^2 - J_{\down\up}\tilP_\up^2}
  {J_c + J_p}\,
\right]
\end{equation}
with $J_c$ and $J_p$ given by Eqs.~(\ref{pair::eq:6}) and
(\ref{pair::eq:7}).

\begin{figure}
\centering
\includegraphics*[width=8cm]{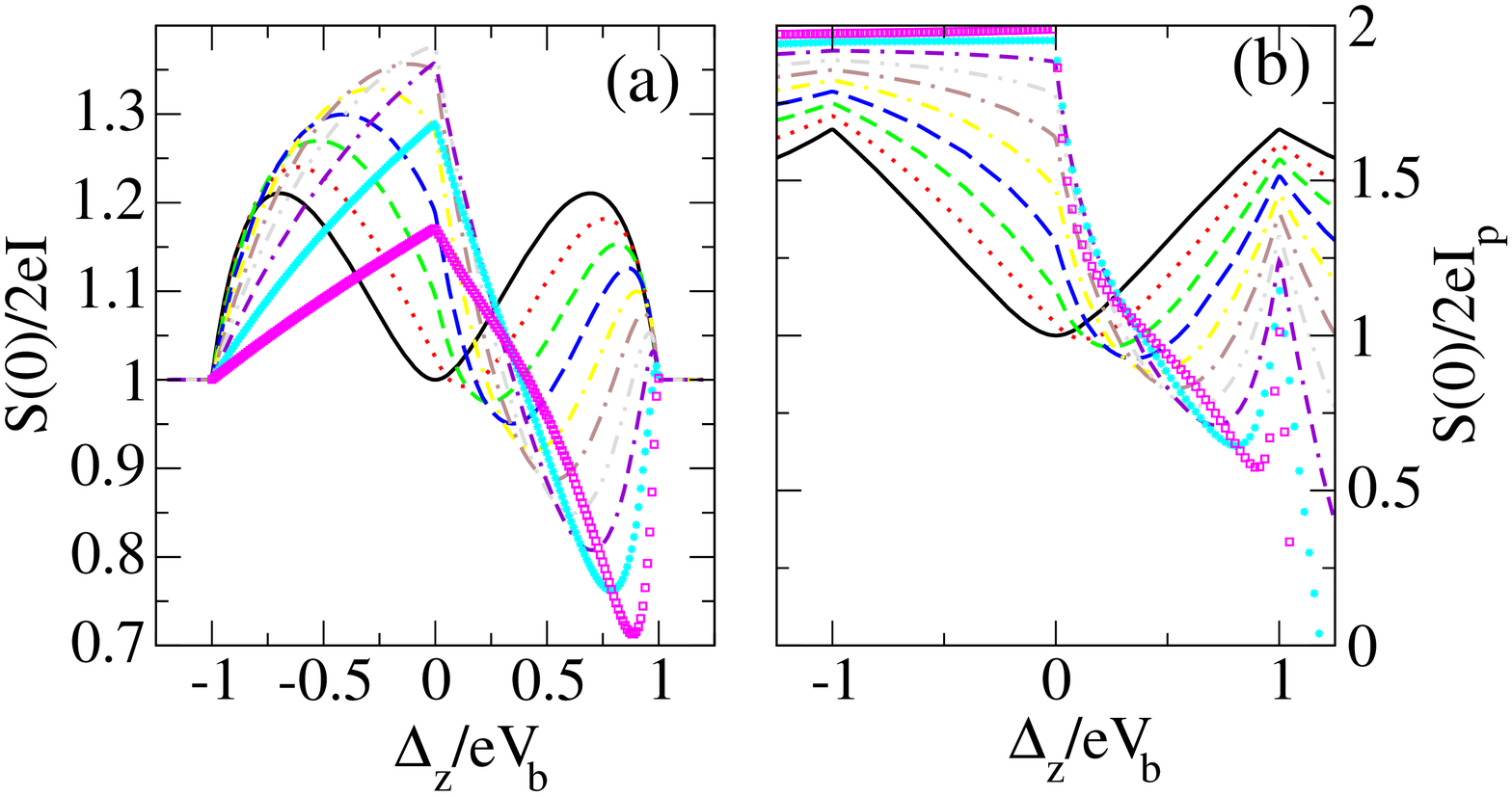}
\caption{(color on-line) (a) Fano factor as a function of $\Delta_z/eV_b$
  for different values of $\alpha$ from $0$ (black solid line) to $0.9$
  (violet empty square) in steps of $0.1$.
  $\alpha\equiv(\Gamma_L-\Gamma_R)/\Gamma$ is the junction asymmetry
  parameter. Here $eV_b=500k_BT$ and $\gamma=1$. (b) Fano factor with
  respect to $I_p$.}
\label{pair::fig:3}
\end{figure}

The shot noise characteristic is summarized in Fig.~\ref{pair::fig:3}.  It
shows several interesting features: (i) The deviation from the
Poissonian shot noise [$S(0)/2eI=1$] comes entirely from the pair
tunneling.  This can be easily understood since the SPC in the
equivalent model is an elastic process, which is known to be
Poissonian~\cite{Sukhorukov01a}.
(ii) For the symmetric junctions ($\Gamma_L=\Gamma_R$), the Fano factor
is exactly 1 at resonance ($\Delta_Z=0$), and increases rapidly with
$|\Delta_Z|$ (up to $|\Delta_Z|\sim eV_b$).
The usual sequential tunneling at resonance across symmetric junctions
gives a Fano factor of $0.5$\cite{shot}. One may be tempted to interpret
it as $0.5\times 2$ with the factor of $2$ for the effective charge $2e$
of pairs.  However, this cannot be justified since the Fano factor is
relative to the total current $I=e(J_c+J_p)$, but not to the
pair-tunneling contribution $I_p=eJ_p$.
For definiteness, let us examine the \emph{pair} Fano factor
$S(0)/2eI_p$, as shown in Fig.~\ref{pair::fig:3} (b).
It shows that the pair tunneling in the system cannot be interpreted as
a tunneling of tightly bound pair objects.\cite{endnote:3}
Instead, we interpret it again in the cotunneling
picture of the equivalent model.  The SFC is an
inelastic process.  At $\Delta_Z=0$, the two channels for the inelastic
cotunneling are equal, and give no additional
fluctuations (the Fano factor of $1$).  As $\Delta_Z$
increases, one channel carries larger current than the other, which
gives additional fluctuations~\cite{Sukhorukov01a}.
(iii) The shot noise turns out to be very sensitive to the junction
asymmetry.  Unlike single-particle sequential tunneling, the
pair-tunneling even gives sub-Poissonian shot noise.
This is also different from the inelastic cotunneling noise in usual
QDs~\cite{Sukhorukov01a}, but this difference comes from the exchange of
the lead indices in the PHT.

\section{Tunneling of Electron-Hole Pairs}
\label{sec:ccdqd}

\begin{figure}
\centering
\includegraphics*[width=80mm]{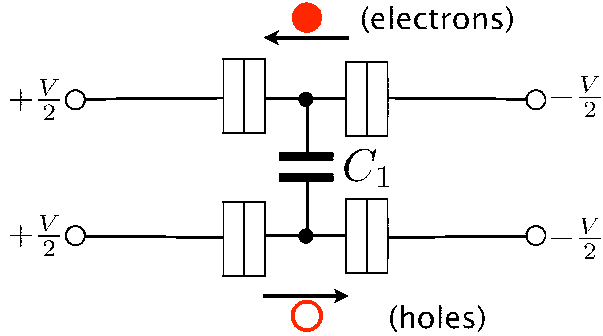}
\caption{(color on-line) A schematic of capacitively-coupled double
  quantum dots.  The majority carriers through QD 1 (upper) are the
  electrons (red filled circle) and those through QD 2 (lower) are the
  holes (red empty circle).  The gate voltages coupled to QDs are not
  shown for simplicity.}
\label{pair::fig:4}
\end{figure}

Now we propose a experimentally feasible set-up to observe the pair
tunneling transport. It consists of capacitively coupled double
quantum dots, where each dot is connected to its own leads via
tunneling junctions and two dots are coupled in parallel
capacitively; see Fig.~\ref{pair::fig:4}. The coupling capacitance
between two dots is denoted by $C_1$, to be distinguished from the
self-capacitance $C_0$ of each quantum dot(QD), which includes the
junction capacitance and the gate capacitance. The tunnel
junctions are specified by the hybridization parameter
$\Gamma_{\ell=\{L,R\},i=\{1,2\}}$ between QD $i$ and lead $\ell$.
Since capacitance is relatively insensitive to the sample
fabrication geometry (unlike exponentially sensitive hybridization),
for simplicity we assume that the capacitances are symmetric over
the upper and lower branches and for the left and right junctions.
The gate voltages are applied oppositely to the two QDs, so that the
gate-induced charges are opposite to each other:
\begin{equation}
n_{g,1} = -n_{g,2} = n_g \,.
\end{equation}
The effective carriers, say, through QD 1 are then electrons while those
through QD 2 are holes, provided that the QDs are made of
semiconductors\cite{Chan}.  When the QDs are ultra-small metallic
grains\cite{Delsing96a,Matters97a}, the ``electrons'' here are the \emph{excess
  electrons} and the holes are the \emph{deficit electrons} with respect to
the mean background charge.

\begin{figure}
\centering
\includegraphics[width=80mm]{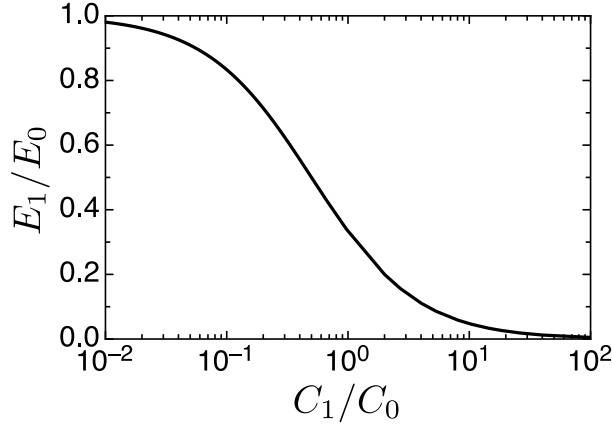}
\caption{The Coulomb energy ratio $E_1/E_0$ versus the capacitance ratio
  $C_1/C_0$. Note that for arbitrary $C_1/C_0$, $E_0$ is always bigger
  than $E_1$, leading to the negative interaction energy $U_C$.}
\label{pair::fig:5}
\end{figure}

The electrostatic energy of the double QD is given by
\begin{equation}
\label{pair::eq:11}
E_C(n_1,n_2) = E_1(n_1+n_2-2n_g)^2 + E_0(n_1-n_2)^2
\end{equation}
where $n_1$ is the number of \emph{excess electrons} on QD 1 while $n_2$
is the number of \emph{excess holes} (the number of \emph{deficit
  electrons}) on QD 2.
In Eq.~(\ref{pair::eq:11}), the two Coulomb energy scales $E_0$ and $E_1$
have been defined by
\begin{equation}
E_0 \equiv \frac{e^2}{4C_0}
\end{equation}
and
\begin{equation}
E_1 \equiv \frac{e^2}{8C_1}\,\frac{1}{1+C_0/2C_1} \,.
\end{equation}
We note that the ratio $E_1/E_0$ is given by
\begin{equation}
\label{pair::eq:12}
\frac{E_1}{E_0} = \frac{C_0/2C_1}{1+C_0/2C_1}
\approx (C_0/2C_1) - (C_0/2C_1)^2 \,,
\end{equation}
and that $E_1$ is always smaller than $E_0$ for arbitrary ratio
$C_1/C_0$ as illustrated ed in Fig.~\ref{pair::fig:5}.  In particular,
for $C_1\gg C_0$, $E_1/E_0$ becomes vanishingly small.

In this large coupling regime, Eq.~(\ref{pair::eq:11}) shows that
the electrostatic energy is very low when there are same number of
electrons and holes in QD1 and QD2, respectively, that is, when
$n_1=n_2$. In other words, it is energetically favorable for this
system to form electron-hole pairs in two dots. Indeed, in the
Coulomb blockade regime ($E_0,E_1\gg \Gamma, k_BT$), if we set the
gate voltage, so that $n_g=1/2$, then one can identify four
lowest-energy charge states of the DQD,
$\ket{n_1,n_2}=\ket{0,0},\ket{1,1},\ket{1,0}$ and $\ket{0,1}$, whose
energies are given by
\begin{equation}
E_C(0,0) = E_C(1,1) = E_1 \,,\quad
E_C(1,0) = E_C(0,1) = E_0
\end{equation}
Therefore, $U_C\equiv 2(E_1-E_0)<0$ plays the role of the negative
interaction energy $U$ of the molecule and $\epsilon_C\equiv(E_0-E_1)>0$
is the counterpart of the energy $\epsilon_d$ of the singly-occupied
states in the molecule.  We emphasize that for arbitrary $C_0/C_1$, one
always has $U_C<0$ and $\epsilon_C>0$; see Eq.~(\ref{pair::eq:12}) and
Fig.~\ref{pair::fig:5}.  Of course, in order for the electron-hole pair
state $\ket{1,1}$ (together with the empty state $\ket{0,0}$) to
dominate the other charge states in the transport, it is required that
\begin{equation}
|U_C|, \epsilon_C \gg \Gamma \,,
\end{equation}
which in turn gives the condition
\begin{equation}
\frac{C_1}{C_0} \gg \frac{1}{2(E_0/\Gamma - 1)} \,.
\end{equation}
In typical experiments on small metal-grain
QDs\cite{Delsing96a,Matters97a,Sohn97a}, $E_0/\Gamma\gtrsim 5$.  It then
suffices that
\begin{equation}
\label{pair::eq:13}
\frac{C_1}{C_0} \gg \frac{1}{10} \,.
\end{equation}
Large capacitive couplings have already been realized experimentally.
$C_1/C_0\approx 5$ in nanoscale metallic grains~\cite{Matters97a}, fully
satisfying the condition~(\ref{pair::eq:13}).  On semiconducting quantum
dots~\cite{Chan} $C_1/C_0\approx 0.28 (0.34)$, satisfying the
condition~(\ref{pair::eq:13}) only marginally and hence reducing
the pair-tunneling contributions slightly from the ideal values.
A very recent measurement of the cross-correlation noise through
CCDQD~\cite{McClure07a}, even though the coupling capacitance in this
experiment was rather small ($C_1/C_0\approx 0.1$) and pair-tunneling
contribution cannot be dominant, suggests that further detailed studies
of the CCDQD are worthwhile.

Given the condition~(\ref{pair::eq:13}) satisfied, the calculations
of the current and noise are exactly the same as in the molecular
case with only one exception.  The exceptional difference comes from
the fact that, for example, $\Gamma_{L,1}$ and $\Gamma_{L,2}$ can be
significantly different in realistic experiments.  These junction
anisotropies change, e.g., the conductance in
Eq.~(\ref{pair::eq:8}) to
\begin{multline}
G = \frac{2e^2}{h}
\frac{\Gamma_{L,1}\Gamma_{R,2}+\Gamma_{L,2}\Gamma_{R,1}}{2} \Bigg[
\frac{U^2}{\epsilon_d^2(\epsilon_d+U)^2}
\frac{\beta(2\epsilon_d+U)}{2\sinh[\beta(2\epsilon_d+U)]}
\Bigg] \\
+ \frac{2e^2}{h}
\frac{\Gamma_{L,1}\Gamma_{R,1}+\Gamma_{L,2}\Gamma_{R,2}}{2} \Bigg[
\frac{1-f(2\epsilon_d+U)}{\epsilon_d^2} +
\frac{f(2\epsilon_d+U)}{(\epsilon_d+U)^2} \Bigg]
\end{multline}
Note that the FC factor $\gamma^2$ does not appear here, and hence for
reasonably symmetric junctions with
\begin{math}
(\Gamma_{L,1}\Gamma_{R,2}+\Gamma_{L,2}\Gamma_{R,1}) \approx
(\Gamma_{L,1}\Gamma_{R,1}+\Gamma_{L,2}\Gamma_{R,2})
\end{math}
the pair-tunneling contribution can be clearly seen on top of the
broad normal-cotunneling background.
The junction anisotropy is expected to enhance the current fluctuations
and hence the super-Poissonian nature of the noise characteristics.

\section{Conclusion}
\label{sec:conclusion}

By exploiting the particle-hole transformation, we studied the pair
tunneling through a single-molecule transistor on an equal footing
to the normal cotunneling.  We have clarified the nature of the pair
tunneling by revealing new features of the shot noise
characteristics as well as reinvestigating the average current. 
We also respected the general observation that pair tunneling is
subject to stronger FC suppression than the normal cotunneling, and
proposed an experimental setup to enhance a visibility of the pair
tunneling.

\begin{acknowledgments}
This work was supported by the SRC program (R11-2000-071), the KRF
Grants (KRF-2005-070-C00055 and KRF-2006-312-C00543), the Second BK21,
the Grant FIS2005 02796 (MEC), and the ``Ram\'on y Cajal'' program.
We thank F.~von Oppen for sending his presentation slides and preprint.
M.-J.H thanks H.-W. Lee for helpful discussions.
\end{acknowledgments}


\begin{thebibliography}{99}
\bibitem{endnote:1} The associated Kondo effect is out of the scope of
this work.
\bibitem{endnote:3} In Fig.~\ref{pair::fig:3}, the noise in the region
$|\Delta_Z|\gg eV_b$ is governed by thermal-assisted tunneling, and is
not ``shot'' noise.
\end{thebibliography}
\end{document}